\documentclass[aps,prd,twocolumn,showpacs,preprintnumbers,
               amsmath,amssymb]{revtex4-2}

\usepackage{amsmath,amssymb}
\usepackage{hyperref}
\usepackage{microtype}
\usepackage{graphicx}
\usepackage{multirow}

\usepackage{amsthm}

\usepackage{booktabs}

\begin{document}

\title{Finite-range Lattice Momentum Operators for Quantum Field Theory}

\author{J.C. Olivier}
\affiliation{Independent Researcher, Australia. Email: jc.olivieriii@gmail.com}

\author{E. Barnard}
\affiliation{ The Faculty of Engineering, North-West University, Potchefstroom, South Africa. Email: etienne.barnard@gmail.com}


\begin{abstract}
We propose a Z-transform framework for the analysis and synthesis 
of finite range lattice momentum operators in quantum field theory. In this  formulation, translation-invariant lattice operators are represented as functions of the complex variable $z$ in the unit circle, allowing their spectral properties to be analyzed using tools from digital signal processing and rational approximation theory. 

Within this 
framework, the fermion doubling problem is reinterpreted as the 
appearance of unwanted zeros of the discrete momentum operator on 
the unit circle --- an aliasing phenomenon in the sense of the 
Nyquist sampling theorem --- and the conditions for ghost suppression 
are expressed as precise constraints on the zero structure of the 
operator's transfer function. An impossibility result, proved in the 
appendix, establishes that no rational function can satisfy all 
required conditions simultaneously, motivating the finite impulse 
response approach developed here.

This reframing naturally suggests a class of finite-range momentum 
operators, constructed by solving a least-squares approximation 
problem in the frequency domain. The resulting finite impulse 
response (FIR) operator approximates the continuum derivative 
across the full Brillouin zone, with ghost suppression achieved 
through the accuracy of the spectral approximation rather than 
through the addition of a symmetry-breaking Wilson term or the 
infinite-range nonlocal SLAC derivative. The ghost mode retains 
an exact zero at the Brillouin zone boundary by symmetry, but 
the spectral width of this zero shrinks as $1/N$ with the number 
of taps $N$, rendering it unable to support coherent ghost 
propagation for any physically realisable wave packet. 
Numerical investigation confirms this mechanism directly: near 
$\theta = \pi$ only plane waves propagate coherently, and these exhibit group velocities far exceeding the speed of light, further distinguishing them from physical low-energy excitations. No ghost wave 
packet solutions exist near $\theta = \pi$.

Numerical simulations of free electron propagation in $1+1$ 
dimensions demonstrate that the proposed FIR operator with 
$N = 30$ taps reproduces the continuum group velocity to within 
$0.06\%$ at $p_0 = \pi/8$ and $0.03\%$ at $p_0 = 0.45\pi$, 
compared to errors of $9\%$ and $78\%$ for the central 
difference operator and $34\%$ and $6\%$ for the Wilson 
operator at the same momenta. Momentum and norm are conserved 
to machine precision throughout. 

The gauge-covariant extension 
of the FIR operator requires \emph{finite-range} parallel transporters, but remains an open problem for future investigation. Additionally  the smooth roll-off of the FIR transfer function near the 
zone boundary avoids the discontinuity that have been reported to be problematic for the SLAC 
derivative in gauge theories.

\end{abstract}


\maketitle

\section{Background and Motivation}



Lattice quantum field theory, pioneered by Wilson~\cite{Wilson}, provides a non-perturbative formulation of quantum field theory by discretizing spacetime with lattice spacing $a$.  The QED Lagrangian itself is uniquely determined by imposing local $U(1)$ gauge invariance on the electron field,
\begin{equation}
\psi(x)\rightarrow e^{i\alpha(x)}\psi(x),
\end{equation}
leading to
\begin{equation}
\mathcal{L} =
\bar{\psi}
(i\gamma^\mu D_\mu-m)
\psi
-
\frac14
F^{\mu\nu}F_{\mu\nu},
\end{equation}
where
\[
D_\mu=\partial_\mu+ieA_\mu.
\]

The interaction term
$-e\bar{\psi}\gamma^\mu\psi A_\mu$,
Maxwell's equations,
and charge conservation
all follow from this single symmetry principle.

Applying the Euler--Lagrange equations yields the coupled QED field equations,
\begin{equation}
(i\gamma^\mu\partial_\mu-m)\psi
=
e\gamma^\mu A_\mu\psi,
\end{equation}
and
\begin{equation}
\partial_\nu F^{\mu\nu}
=
e\bar{\psi}\gamma^\mu\psi
\equiv
J^\mu.
\end{equation}
A direct numerical solution of these coupled equations on the lattice, given suitable initial conditions, will in principle reproduce the predictions of QED.

A central motivation for the present work is the tension between
spectral accuracy and locality in lattice fermion formulations. The
SLAC derivative provides an instructive example \cite{Drell}. By reproducing the
continuum momentum relation across the Brillouin zone, it avoids the
additional low-energy zero associated with the conventional central
difference operator. However, this spectral accuracy is obtained at
the cost of an operator with infinite spatial range. In the presence
of gauge interactions, the coupling of fermion fields at arbitrarily
separated lattice sites requires corresponding gauge transport over
arbitrarily long distances. The resulting nonlocal gauge structure has
been identified as a source of difficulty for SLAC fermions in
interacting gauge theories \cite{Duitsers}.

The guiding assumption of the present work is therefore that a useful
lattice momentum operator should have a strictly finite coupling
range. Such an operator requires only \emph{finite-range} gauge transport
when gauge interactions are introduced. Hence,  its implementation
is local in the spatial sense, and potentially avoids the difficulties
associated with the infinite-range transport required by the SLAC
construction. We emphasize, however, that this conjecture is not
established here: the present work considers free fermion propagation,
and the behavior of the proposed operators in fully gauge-interacting
theories remains an important subject for future investigation.

The question then becomes how to construct a finite-range operator
while retaining as much as possible of the desirable spectral
properties of the continuum derivative. We find that the
$Z$-transform provides a particularly natural framework for this
purpose. Translation invariance converts the lattice convolution
operator into a function of a single complex variable, while
anti-Hermiticity, locality, the continuum limit, and the structure of
the zeros become explicit mathematical constraints on that function.
The construction can consequently be formulated as a constrained
approximation problem, allowing the coefficients of a finite impulse
response (FIR) operator to be obtained systematically rather than
chosen ad hoc.

The same formulation also provides a transparent way of understanding
the underlying no-go constraint. The periodicity of the $Z$-plane
representation makes the additional zero at the Brillouin-zone
boundary unavoidable under the assumptions of the construction. Thus,
the proposed approach does not claim to evade the
Nielsen--Ninomiya obstruction \cite{Nielsen1,Nielsen2}. Instead, it asks a different question:
\emph{can the unavoidable zero be confined to such a narrow spectral region that it cannot support a localized, coherently propagating ghost excitation?} Our numerical results indicate that it
can. As the FIR order is increased, the dispersion approaches the
continuum relation over an increasingly large fraction of the
Brillouin zone, while a ghost excitation near the residual boundary
zero must become increasingly narrow in momentum space in order to
propagate coherently. Such a state approaches a
plane wave even for a moderate coupling range, and is completely delocalized in position space.

The resulting picture is therefore a compromise between the two
extremes. The SLAC construction achieves an essentially exact
continuum dispersion at the cost of infinite spatial range, whereas
the present approach retains a strictly finite coupling range and
accepts a controlled spectral approximation. The central question is
whether this trade-off provides a practically useful lattice fermion
formulation. The numerical results presented below provide evidence
that high spectral accuracy and suppression of coherent ghost
propagation can be achieved simultaneously with strictly finite
coupling range. Whether the resulting operators retain these
advantages in the presence of gauge interactions remains an important
question for future work.


\section{The Fermion Doubling Problem and Aliasing}

The principal obstacle to the naive lattice discretization of the Dirac equation is the fermion doubling problem \cite{Wilson,Nielsen1,Montvay}. When the electron field is sampled on a lattice with spacing $a$, the momentum-space representation becomes periodic with period $2\pi/a$, restricting physical momenta to the first Brillouin zone,
\begin{equation}
k\in
\left[
-\frac{\pi}{a},
\,
\frac{\pi}{a}
\right].
\end{equation}
Using the standard central-difference approximation,
\begin{equation}
\partial_x\psi(x)
\approx
\frac{\psi(x+a)-\psi(x-a)}{2a},
\end{equation}
the free Dirac operator becomes
\begin{equation}
S^{-1}(k)
=
\gamma^\mu
\frac{\sin(k_\mu a)}{a}
+
m.
\end{equation}

In the Hamiltonian formulation, where time remains continuous 
and only the spatial direction is discretized, the central 
difference operator $\partial_x \approx (\psi(x+a) - 
\psi(x-a))/(2a)$ has transfer function $\sin(k_1 a)/a$, 
which vanishes at both $k_1 = 0$ and $k_1 = \pi/a$. This 
produces $2^{d_s}$ fermion species in $d_s$ spatial dimensions 
rather than one, where $d_s$ is the number of discretized 
spatial dimensions.

From the perspective of digital signal processing, this behaviour can be interpreted as an aliasing phenomenon. Sampling a continuous field at spacing $a$ produces a periodic momentum spectrum with period $2\pi/a$, exactly as sampling a continuous-time signal produces periodic spectral replicas in the discrete-time Fourier transform \cite{Oppenheim}. The additional fermion species may therefore be viewed as lattice analogues of aliased spectral images. This interpretation does not alter the underlying physics; rather, it provides an alternative mathematical language in which the doubling phenomenon can be understood.

Wilson's original solution introduces an additional second-order difference operator (the Wilson term), assigning masses of order $1/a$ to the unwanted species so that they decouple in the continuum limit \cite{Wilson}. The price is explicit breaking of chiral symmetry. Subsequent developments—including staggered fermions, domain-wall fermions and overlap fermions \cite{Neuberger1998} address different aspects of this compromise, but each introduces additional mathematical or computational complexity.

The so-called SLAC fermion operator \cite{Drell} is of specific relevance to the proposed Z-transform formulation of the lattice momentum operator presented in the next section. For the one-dimensional periodic lattice, the ideal SLAC momentum operator represented on the unit circle is given by 
\begin{equation}
    D_{\mathrm{SLAC}}(\theta) = i\theta,
    \qquad -\pi < \theta < \pi.
\end{equation}
Fourier transforming this expression gives a position-space kernel
that has an infinite $1/n$-type tail. In other words,
\begin{equation}
    h_n \sim \frac{(-1)^n}{n}.
\end{equation}
This shows that the coupling is infinite, presenting serious difficulty when Gauge fields (interactions) are included \cite{Duitsers}.  Once distant lattice sites are coupled,
\begin{equation}
    \psi(n) \longleftrightarrow \psi(m),
\end{equation}
gauge covariance requires the fields at the two sites to be connected
by a gauge transporter,
\begin{equation}
    \psi(n) \longleftrightarrow
    U(n\rightarrow m)\,\psi(m).
\end{equation}
This introduces a nonlocal gauge structure into the theory due to the infinite range position-space kernel.

There is a second issue with the SLAC fermion operator, and that has to do with the discontinuity at $\theta = \pi$ \cite{Duitsers}.  The authors in \cite{Duitsers} conclude that the combination of non-locality and the zone-edge singularity are the fundamental difficulty for SLAC fermions in gauge theories.

\section{The Z-Transform Reformulation for Lattice Operators with Finite Range}

The central idea of the present work is to reformulate the lattice momentum operator using the language of the Z-transform. In this setting, the search for an admissible lattice momentum operator becomes a problem in rational approximation on the unit circle.

Table~\ref{tab:correspondence} summarises the correspondence between concepts in lattice quantum field theory and linear systems theory.
\begin{table}[t]
\centering
\small
\begin{tabular}{ll}
\hline
\textbf{Quantum field theory} & \textbf{Linear systems theory}\\
\hline
Translation operator & Delay operator\\
Momentum & Phase on the unit circle\\
Brillouin zone & Nyquist interval\\
Fermion doubling & Aliasing image\\
Derivative operator & Digital differentiator\\
Locality & Impulse response, finite support\\
\hline
\end{tabular}
\caption{Correspondence between lattice QFT and linear systems theory  concepts.}
\label{tab:correspondence}
\end{table}

The bilateral Z-transform of a lattice field $\psi(n)$ is
\begin{equation}
\Psi(z) = \sum_{n=-\infty}^{\infty}
\psi(n)z^{-n}.
\end{equation}
Within this representation, a translation by one lattice spacing corresponds simply to multiplication by $z$,
\begin{equation}
U(a)
\longleftrightarrow
z,
\end{equation}
where $U(a)$ denotes the unitary translation operator.

\subsection{Momentum eigenvalues on the unit circle}

Since $U(a)$ is unitary, its eigenvalues necessarily lie on the unit circle,
\[
|z|=1.
\]
Hence it follows that 
\[
z=e^{ipa/\hbar}
\]
maps the physical momentum interval onto the unit circle,
\begin{equation}
z=e^{ipa/\hbar},
\qquad
p\in
\left[
-\frac{\pi\hbar}{a},
\,
\frac{\pi\hbar}{a}
\right].
\end{equation}
The origin corresponds to
\[
z=1,
\]
while the Nyquist momentum corresponds to
\[
z=-1.
\]

This restriction of the translation spectrum to the unit circle is not peculiar to lattice discretization. Rather, it follows from the unitary representation of continuous spacetime symmetries on Hilbert space, as required by Wigner's theorem (see, for example, Weinberg~\cite{Weinberg}).

Consequently, the physical momentum spectrum is naturally represented as points on the unit circle, making the Z-transform the appropriate mathematical setting for studying discrete finite range momentum operators.

\subsection{Fermion doubling in the Z-domain}

The naive lattice momentum operator corresponding to the central difference has Z-transform
\begin{equation}
\hat p_{\rm naive}
\longleftrightarrow
\frac{\hbar}{ia}
\frac{z-z^{-1}}{2}
=
\frac{\hbar}{ia}
\frac{z^2-1}{2z}.
\end{equation}
The numerator contains the factor
\[
z^2-1=(z-1)(z+1),
\]
revealing two distinct zeros on the unit circle,
\[
z=1
\qquad\text{and}\qquad
z=-1.
\]
The first corresponds to the physical zero-momentum state, while the second corresponds to the unwanted fermion doubler at the edge of the Brillouin zone.

Viewed from the perspective of digital filter theory, the doubling problem is therefore equivalent to constructing a differentiator whose transfer function possesses an unwanted additional zero on the unit circle. This observation motivates the reformulation proposed here: rather than viewing fermion doubling solely as a lattice field theory problem, one may instead regard it as a problem in the design of rational differentiators.

A natural objective is therefore to seek a lattice momentum operator whose Z-domain representation possesses exactly one zero on the unit circle, namely at
\[
z=1.
\]
Such an operator would represent a single fermion species, eliminate the spurious zero at the Brillouin-zone boundary, preserve the required Hermiticity properties of the continuum momentum operator, and recover the continuum derivative as $a\rightarrow0$.

The continuum limit has a simple interpretation in this framework. As the lattice spacing tends to zero, the Brillouin zone expands to cover the real momentum axis while the unit-circle representation converges to the continuum spectrum. The problem therefore reduces to finding a discrete momentum operator whose finite-lattice representation converges appropriately to its continuum counterpart.

\subsection{The Anti-Hermitian Constraint}

However, there is another fundamental requirement. In the continuum,
\begin{equation}
\int
\bar\psi
(\partial_x\psi)\,dx
=
-
\int
(\partial_x\bar\psi)\psi\,dx,
\end{equation}
so that
\[
(\partial_x)^\dagger
=
-\partial_x.
\]
Thus the derivative operator is anti-Hermitian, ensuring that the momentum operator
\[
\hat p
=
-i\hbar\partial_x
\]
is Hermitian. Hermiticity of the Hamiltonian guarantees unitary time evolution and conservation of probability.  Expressed in the Z-domain, anti-Hermiticity requires
\begin{equation}
D(e^{i\theta})^*
=
-
D(e^{-i\theta}),
\end{equation}
for every point on the unit circle.
Equivalently,
\[
D(e^{i\theta})
=
i\,f(\theta),
\]
where $f(\theta)$ is a real-valued odd function.


\section{The Mathematical Existence Problem}

The preceding discussion naturally leads to the following mathematical problem: determine whether there exists a rational function $D(z)$ satisfying the following four conditions simultaneously.
\begin{enumerate}
\item
\textbf{Unique physical zero.}
The function possesses exactly one zero on the unit circle, located at
\[ z=1. \]
\item
\textbf{Anti-Hermiticity.}
On the unit circle,
\[
D(e^{i\theta})
=
i\,f(\theta),
\]
where $f(\theta)$ is real-valued, odd, and non-zero for
\[
\theta
\in
(-\pi,\pi)
\setminus
\{0\}.
\]
\item
\textbf{Locality.}
$D(z)$ is a rational function of finite degree, corresponding to finite-range couplings in position space.
\item
\textbf{Correct continuum limit.}
As \[ a\rightarrow0, \] the operator converges to the continuum momentum operator, \[ f(\theta) \rightarrow \frac{\theta}{a}. \]
\end{enumerate}





The existence problem provided above can be shown not to have an exact solution, and the proof is provided in the Appendix. The impossibility of satisfying all the desired properties exactly is consistent with the known no-go theorems of lattice fermions \cite{Nielsen1,Nielsen2}, but the Z domain  perspective and proof provided here show that the Z-domain perspective simplifies the analysis. 

\subsection{Rational Approximation and Finite Range Lattice Operators}

The impossibility of satisfying all of the desired properties exactly
within the class of rational momentum operators shifts the problem from
one of exact construction to one of constrained approximation.
The Z-transform formulation naturally suggests constructing
families of rational operators whose spectral properties approximate
the desired momentum operator while maintaining locality in an
approximate sense.


First of all, we now focus on finite range operators, that is a linear filter with a Finite Impulse Response (FIR).  The convolution form of the momentum operator requires careful
attention to the symmetry of the impulse response $h(n)$. A one-sided
(causal) impulse response is generally incompatible with
anti-Hermiticity, since its Z-transform is generally not purely
imaginary and odd on the unit circle.

For a translation-invariant convolution operator, anti-Hermiticity is equivalent to the real impulse response satisfying
\begin{equation}
    h(-n)=-h(n), \qquad h(0)=0,
\end{equation}
so that the momentum operator takes the convolution form
\begin{equation}
    (\hat p\psi)(n)
    =
    \sum_{k=1}^{N}
    h(k)\,\bigl(\psi(n-k)-\psi(n+k)\bigr).
\end{equation}
The antisymmetry of $h$ ensures that left and right neighboring lattice sites contribute symmetrically. This two-sided structure is not an additional requirement but a direct consequence of anti-Hermiticity. 

To see why the proposed finite-range operator and its construction may be more amenable to incorporating gauge fields than the SLAC operator, it is useful to consider how the two operators are implemented. The SLAC derivative
is naturally evaluated in momentum space, typically by transforming
the field with an FFT, multiplying by $ik$, and transforming back.
The FFT itself provides no local point at which the gauge field can be
inserted: the resulting derivative couples every lattice site to every
other site. A gauge-covariant implementation therefore requires
parallel transport of the field between arbitrarily separated sites,
with the corresponding gauge links inserted into the nonlocal
operator. In dimensions greater than one, the choice of transport
path introduces additional structure because parallel transport is
generally path dependent \cite{Karsten}.

The FIR operator, by contrast, is defined directly in position space
and has strictly finite support. It couples site $n$ only to the
sites $n\pm k$, with $k=1,\ldots,N$. A natural gauge-covariant
extension is therefore obtained by replacing each displacement by
the corresponding ordered product of gauge links,
\begin{equation}
    D_{\mathrm{FIR}}^{\mathrm{cov}}\psi(n)
    =
    \sum_{k=1}^{N} h(k)
    \left[
        U_n^{(k)}\psi(n+k)
        -
        U_n^{(-k)}\psi(n-k)
    \right],
\end{equation}
where, for example,
\begin{equation}
    U_n^{(k)}
    =
    \prod_{j=0}^{k-1} U_{n+j}.
\end{equation}
In $1+1$ dimensions there is a unique straight lattice path between
the two sites, making this construction particularly transparent.
In higher dimensions, different paths can in general produce
different parallel transporters, and the corresponding construction
requires an explicit choice of path or path averaging. Whether the
finite-range FIR construction retains its favourable properties in
the presence of dynamical gauge fields, and how its finite transport
range affects the resulting interacting theory, are questions left
for future investigation.

\subsection{Transfer function of the FIR operator}

The transfer function of any antisymmetric FIR filter with real
coefficients $h(n)$ is
\begin{equation}
    H(e^{i\theta})
    =
    \sum_{n=1}^{N} h(n)\bigl(e^{-in\theta}-e^{in\theta}\bigr)
    =
    -2i\sum_{n=1}^{N} h(n)\sin(n\theta).
\end{equation}
Since $h(n)$ are real and $\sin(n\theta)$ is real for all real
$\theta$, the transfer function is {purely imaginary} for all
$\theta$ and all choices of $h(n)$. Writing
$H(e^{i\theta})=i\,f(\theta)$, the function
\begin{equation} \label{nul_by_pi}
    f(\theta) = 2\sum_{n=1}^{N} h(n)\sin(n\theta)
\end{equation}
is real-valued and odd in $\theta$ by construction. Three properties
therefore hold exactly, independent of the specific choice of
coefficients:
\begin{enumerate}
    \item \textbf{Anti-Hermiticity.}
    $\operatorname{Re}[H(e^{i\theta})]=0$ for all $\theta$,
    so the operator $i\hat{p}$ is Hermitian and time evolution
    is unitary.
    \item \textbf{Antisymmetry.}
    $H(e^{-i\theta})=-H(e^{i\theta})$, so the operator couples
    left and right neighbours symmetrically.
    \item \textbf{Zero at the origin.}
    $H(e^{i\cdot 0})=0$ exactly, since $\sin(0)=0$ for every
    term in the sum.
\end{enumerate}
These properties are structural consequences of the antisymmetric
FIR form and require no additional enforcement or numerical
verification. 

\begin{figure}[t]
    \centering \includegraphics[width=1.1\columnwidth]{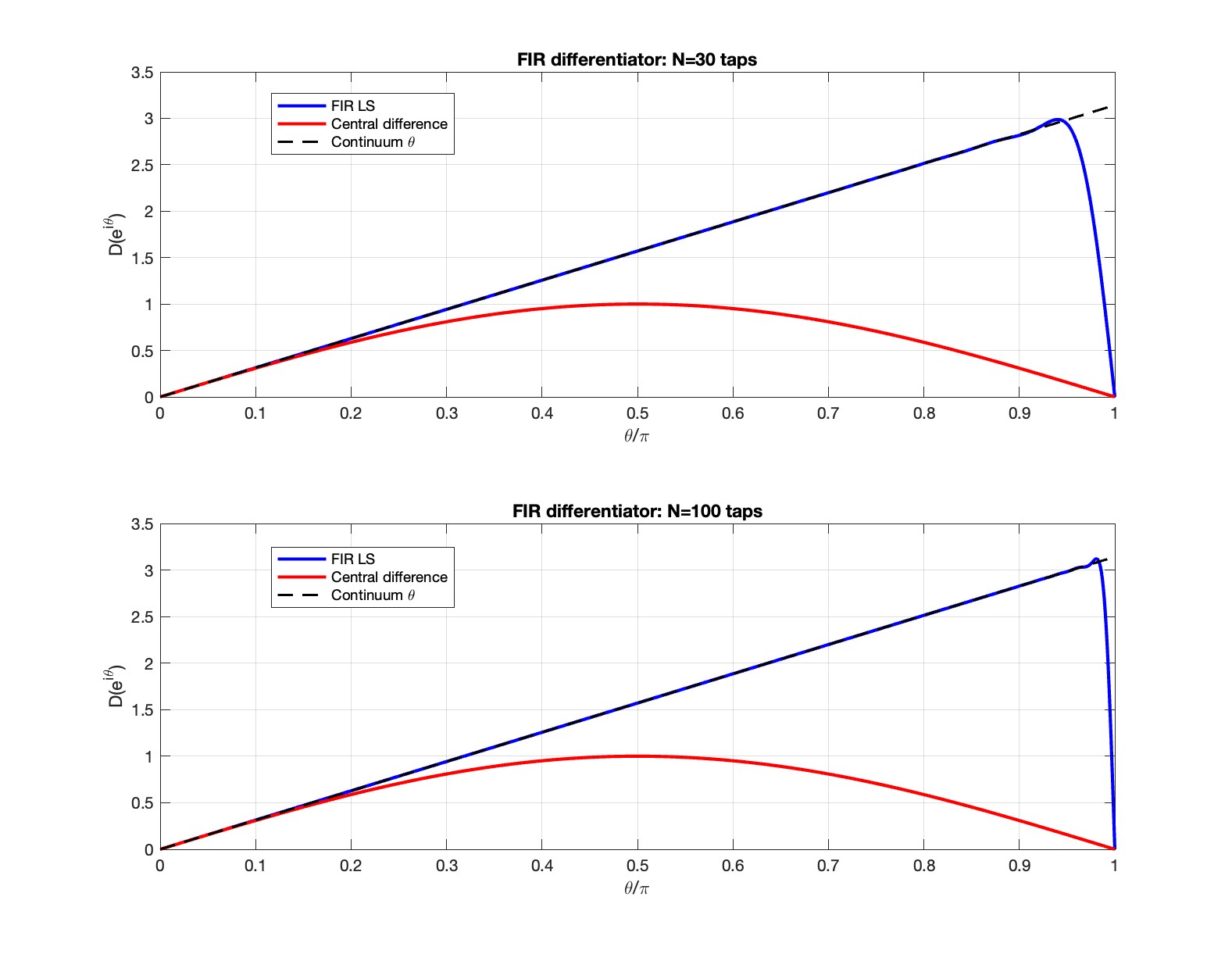} 
    \caption{The dispersion functions of finite range coupling FIR operators. }
    \label{dispersie}
\end{figure}

Figure \ref{dispersie} shows the dispersion curve for two finite range filters. It is clear that an operator that couples a few as $30$ lattice points already has a small low energy spectral gap where $\theta \approx \pi$.  Admittedly, this aspect of the proposed operator must be studied in future work, especially  when gauge fields (interaction) are present. 

\subsection{The approximation problem}

The continuum momentum operator corresponds to the ideal
differentiator response
\begin{equation}
    H_{\mathrm{cont}}(e^{i\theta}) = i\theta,
    \qquad \theta\in[-\pi,\pi].
\end{equation}
Since $H(e^{i\theta})$ is purely imaginary by construction, the
approximation problem reduces entirely to the real condition
\begin{equation}
    f(\theta) = 2\sum_{n=1}^{N} h(n)\sin(n\theta) \approx \theta,
    \qquad \theta\in[0,\pi].
\end{equation}
This is a \emph{real} least-squares problem: find the $N$ real
coefficients $h(1),\ldots,h(N)$ such that the truncated sine series
best approximates the linear function $\theta$ over the Brillouin zone.   Sampling at $M\gg N$ equally spaced points
$\theta_j = j\pi/M$, $j=1,\ldots,M$, the problem takes the matrix
form
\begin{equation} \label{LS_oplossing}
    \min_{\mathbf{h}\in\mathbb{R}^N}
    \bigl\|\mathbf{A}\mathbf{h} - \mathbf{b}\bigr\|^2,
    \qquad
    A_{jn} = \sin(n\theta_j),
    \qquad
    b_j = \frac{\theta_j}{2},
\end{equation}
with solution $\mathbf{h}=\mathbf{A}^{\dagger}\mathbf{b}$ via the pseudoinverse. To produce smooth approximations,  $\theta = \pi$ was excluded in (\ref{LS_oplossing}).

\section{Numerical Validation: Free Electron Propagation}

\subsection{Experimental Design}

To validate the proposed FIR momentum operator and compare it against 
the central difference and Wilson operators, we simulate the free 
propagation of a single electron wave packet across a one-dimensional 
lattice of $N_x = 512$ sites with spacing $a = 1$ (natural units 
$\hbar = c = 1$). The electron mass is $m = 0.5$ and the initial 
state is a Gaussian wave packet,
\begin{equation}
    \psi(n, 0) = \mathcal{N}\,
    e^{-(n - n_0)^2 / 2\sigma^2}\,
    e^{ip_0 n a},
\end{equation}
with width $\sigma = 40$ lattice sites and initial position 
$n_0 = N_x/5$. The spinor is initialised as a right-moving 
eigenstate of the massive Dirac equation. Time evolution uses 
the fourth-order Runge--Kutta (RK4) scheme with time step 
$\Delta t = 0.05$ for a total time $T = 150$ lattice units.

Two initial momenta are considered:
\begin{enumerate}
    \item $p_0 = \pi/8$: well inside the Brillouin 
    zone, where all operators are expected to perform reasonably.
    \item $p_0 = 0.45 \pi$: This is where the central difference operator is known to fail and differences between operators become pronounced.
\end{enumerate}

Three momentum operators are compared: the central difference, the 
Wilson operator, and the proposed FIR operator, and in this case we selected  $N=30$ and $N=100$ tap FIR operators.

\subsection{Theoretical Group Velocities}

For a free electron the group velocity determines the speed at which 
the wave packet centre of mass translates across the lattice. Each 
discrete operator has its own dispersion relation, and hence its 
own predicted group velocity, which differs from the continuum value.

\subsubsection{Continuum}

In the continuum, the massive Dirac dispersion relation is
\begin{equation}
    E_{\mathrm{cont}}(p) = \sqrt{p^2 + m^2},
\end{equation}
giving the group velocity
\begin{equation}
    v_g^{\mathrm{cont}} = \frac{dE}{dp} = \frac{p_0}{E_{\mathrm{cont}}}.
\end{equation}

\subsubsection{Central Difference}

The central difference operator replaces the continuum momentum $p$ 
by $\sin(p_0 a)/a$ in the dispersion relation,
\begin{equation}
    E_{\mathrm{CD}}(p_0) = \sqrt{\frac{\sin^2(p_0 a)}{a^2} + m^2},
\end{equation}
giving the group velocity
\begin{equation}
    v_g^{\mathrm{CD}} = \frac{\sin(p_0 a)\cos(p_0 a)}{a\, E_{\mathrm{CD}}}.
\end{equation}
Near $p_0 = 0$ this agrees with the continuum. Near $p_0 = \pi/(2a)$, 
however, $\cos(p_0 a) \to 0$ and the group velocity vanishes --- the 
wave packet stalls. At $p_0 = \pi/a$ the operator has a second zero, 
producing the fermion doubler with zero energy.

\subsubsection{Wilson Operator}

The Wilson operator adds a discrete Laplacian term to the central 
difference,
\begin{multline}
       D_W \psi(n) = \frac{\psi(n+1) - \psi(n-1)}{2a} - \\ \frac{r_W a}{2}\,\frac{\psi(n+1) - 2\psi(n) + \psi(n-1)}{a^2},
\end{multline}
where $r_W = 1$ is the standard Wilson parameter. The Laplacian term 
is real-valued and acts as a momentum-dependent mass correction. The 
full dispersion relation is
\begin{equation}
    E_W(p_0) = \sqrt{
        \frac{\sin^2(p_0 a)}{a^2}
        + \left(m + \frac{r_W(1 - \cos(p_0 a))}{a}\right)^2
    }.
\end{equation}
At $p_0 = 0$ the mass is $m$, recovering the correct physical mass. 
At $p_0 = \pi/a$ the mass becomes $m + 2r_W/a$, which diverges as 
$a \to 0$, suppressing the ghost mode. 

The correct implementation of the Wilson operator in the Dirac 
equation requires separating the derivative part (purely imaginary, 
entering the off-diagonal spinor coupling) from the mass correction 
part (real, entering the diagonal mass term):
\begin{align}
    \frac{d\psi_{\uparrow}}{dt} &= 
    -D_{\mathrm{CD}}\psi_{\downarrow} 
    - i\left(m + M_W\right)\psi_{\uparrow}, \\
    \frac{d\psi_{\downarrow}}{dt} &= 
    -D_{\mathrm{CD}}\psi_{\uparrow} 
    + i\left(m + M_W\right)\psi_{\downarrow},
\end{align}
where $D_{\mathrm{CD}}$ is the central difference and $M_W(n) = 
-(r_W a/2)\nabla^2$ is the Wilson mass correction. This separation 
is essential: applying the full complex Wilson operator directly in 
the Dirac equation introduces a non-Hermitian contribution that 
violates norm conservation and leads to numerical instability.

\subsubsection{FIR Operator}

The proposed FIR operator with $N$ taps is designed by solving the 
least-squares problem
\begin{equation}
    \min_{\mathbf{h} \in \mathbb{R}^N}
    \left\| \mathbf{A}\mathbf{h} - \mathbf{b} \right\|^2,
    \qquad
    A_{jn} = \sin(n\theta_j),
    \qquad
    b_j = \frac{\theta_j}{2},
\end{equation}
sampled at $M \gg N$ equally spaced points $\theta_j = j\pi/M$.  

The resulting transfer function approximates the ideal continuum 
differentiator $H(e^{i\theta}) = i\theta$ over the full Brillouin 
zone. The dispersion relation is
\begin{equation}
    E_{\mathrm{FIR}}(p_0) = \sqrt{D_{\mathrm{FIR}}(p_0)^2 + m^2}
\end{equation}
where $ D_{\mathrm{FIR}}(p_0) = 2\sum_{n=1}^{N} h(n)\sin(np_0 a).$
Because the least-squares problem fits $D_{\mathrm{FIR}}$ to the 
continuum derivative $\theta/a$, the 
FIR dispersion relation tracks the continuum closely throughout the 
Brillouin zone, as shown in Figure~\ref{dispersie}.

As with the central difference, the FIR operator 
has an exact zero at $\theta = \pi$, which follows directly from 
the antisymmetric basis functions:
\begin{equation}
    D_{\mathrm{FIR}}(\pi) 
    = 2\sum_{n=1}^{N} h(n)\sin(n\pi) = 0.
\end{equation}
The FIR operator yields zero only in a narrow region near $\theta = \pi$. The spectral width of this region is of order $1/N$ in units of $\pi/a$, shrinking as the number of taps 
increases. 

In this sense the ghost mode at $\theta \approx \pi$ is not eliminated, but rather strongly suppressed.  A ghost mode confined to such a narrow spectral window cannot form a coherent propagating wave packet of 
finite spatial extent, since any physically realisable packet 
has momentum spread $\Delta p \sim 1/\sigma \gg 1/(Na)$ for 
reasonable packet widths $\sigma$. Ghost suppression therefore 
improves systematically with $N$ and is already effective for 
moderate tap counts ($30$ taps), which was confirmed by numerical propagation  experiments. Unlike the Wilson 
operator, which suppresses the ghost by introducing an explicit 
momentum-dependent mass correction that breaks chiral symmetry, the 
FIR operator achieves ghost suppression purely through the 
accuracy of its approximation to the continuum derivative --- 
no symmetry is broken and no additional parameter is introduced.

The three structural properties of the antisymmetric FIR operator --- 
purely imaginary transfer function, odd symmetry, and exact zero at 
$\theta = 0$ --- are guaranteed by construction and do not require 
numerical enforcement. This means the FIR operator slots directly 
into the off-diagonal coupling of the Dirac equation without the 
mass correction complication that arises with the Wilson operator.

\subsection{Numerical Results}

Table~\ref{tab:propagation} summarises the measured group velocities, 
momentum conservation, and norm conservation for six runs. The 
measured group velocity is obtained by fitting a straight line to the 
centre of mass trajectory in the final 40\% of the simulation, after 
any initial transients have decayed.

\begin{table}[t]
\centering
\scriptsize
\setlength{\tabcolsep}{2.5pt}
\begin{tabular}{@{}llccccc@{}}
\hline
$p_0$ & Operator & $v_{\rm meas}$ & $v_{\rm exp}$ & $v_{\rm cont}$ & Err. & $|\Delta p|/p_0$ \\
\hline
$\pi/8$ & Central diff & 0.5611 & 0.5615 & 0.6177 & 9.15\% & $<10^{-6}\%$ \\
& Wilson & 0.8292 & 0.8300 & 0.6177 & 34.25\% & $<10^{-6}\%$ \\
& FIR $N=30$ & \textbf{0.6173} & \textbf{0.6176} & \textbf{0.6177} & \textbf{0.06\%} & $<10^{-6}\%$ \\
& FIR $N=100$ & \textbf{0.6174} & \textbf{0.6177} & \textbf{0.6177} & \textbf{0.05\%} & $<10^{-6}\%$ \\
\hline
$0.45\pi$ & Central diff & 0.2076 & 0.2076 & 0.9428 & 77.98\% & $<0.001\%$ \\
& Wilson & 0.8884 & 0.8885 & 0.9428 & 5.77\% & $0.001\%$ \\
& FIR $N=30$ & \textbf{0.9425} & \textbf{0.9425} & \textbf{0.9428} & \textbf{0.03\%} & $<0.001\%$ \\
& FIR $N=100$ & \textbf{0.9426} & \textbf{0.9426} & \textbf{0.9428} & \textbf{0.00\%} & $<0.001\%$ \\
\hline
\end{tabular}
\caption{Free electron propagation results for four momentum operators at two initial momenta. $v_{\rm meas}$ is the group velocity measured from the centre-of-mass trajectory; $v_{\rm exp}$ is the prediction from the corresponding lattice dispersion relation; and $v_{\rm cont}$ is the continuum prediction. The error is the percentage deviation of the measured velocity from the continuum value. $|\Delta p|/p_0$ measures fractional momentum change over the simulation. All operators are initialised with wave packets that are exact eigenstates of their respective discrete Dirac operators.}
\label{tab:propagation}
\end{table}

\subsection{Discussion}

\emph{Momentum conservation} is excellent for all three operators 
at both momenta, with fractional changes below $10^{-5}$ throughout. 
This confirms that the numerical errors reside in the dispersion 
relation (phase velocity) rather than in the norm or momentum 
conservation (amplitude). The FIR operator therefore achieves 
superior dispersion accuracy without sacrificing the conservation 
properties that make the simulation physically reliable.

\emph{The role of correct initialisation} deserves emphasis. 
Each operator defines its own dispersion relation and hence its 
own set of eigenstates. Initialising the wave packet as an eigenstate of the operator under test  is essential for a clean measurement of 
the group velocity.

These results confirm that the proposed FIR least-squares momentum 
operator provides a substantially more accurate discretisation of 
the continuum derivative than either the central difference or 
Wilson operator. 

The performance improvement comes at the cost of increased 
spatial coupling range --- $2N+1 = 61$ lattice sites for $N=30$ 
taps --- but this coupling remains strictly finite, the coefficients 
are precomputed once, and the operator is applied as a single 
convolution at each time step. 

\subsection{Ghost-mode packet propagation}

Although the finite-$N$ FIR operator retains the symmetry-enforced zero at
the Brillouin-zone boundary, numerical propagation experiments show that
this zero does not necessarily support a localized, coherently propagating
ghost wave packet. For example, for $N=30$, the dispersion relation varies
rapidly in a narrow region approaching $\theta=\pi$. Consequently, a wave
packet centered in this region with finite momentum bandwidth contains
components having substantially different group velocities,
\begin{equation}
    v_g(\theta) = \frac{dE(\theta)}{d\theta},
\end{equation}
and therefore undergoes strong dispersion rather than coherent
translation. A sufficiently narrow momentum distribution can still
propagate coherently near the boundary, but such a state approaches a
plane wave and is correspondingly delocalized in position space. Thus,
the finite-$N$ operator does not eliminate the boundary zero; rather, it
restricts the momentum interval over which a low-energy ghost mode can
propagate coherently. As $N$ is increased, this interval becomes
progressively narrower, while the dispersion over the remainder of the
Brillouin zone approaches the continuum relation. The numerical results
therefore suggest that increasing the number of FIR taps progressively
suppresses the ability of the boundary zero to support a localized ghost
excitation, while remaining consistent with the underlying no-go
constraint.

An additional feature of the residual ghost sector is that its
low-energy states, which in the coherent limit are plane waves, occur
in a region of anomalously large group velocity. This observation
suggests that bounded group velocity could provide an additional
constraint in the synthesis of finite-range lattice operators. We do
not pursue such a constraint here, but regard it as a potentially
useful direction for future work.

\subsection{Short-range FIR operators: $10$ taps}

\begin{figure}[t]
    \centering \includegraphics[width=1.1\columnwidth]{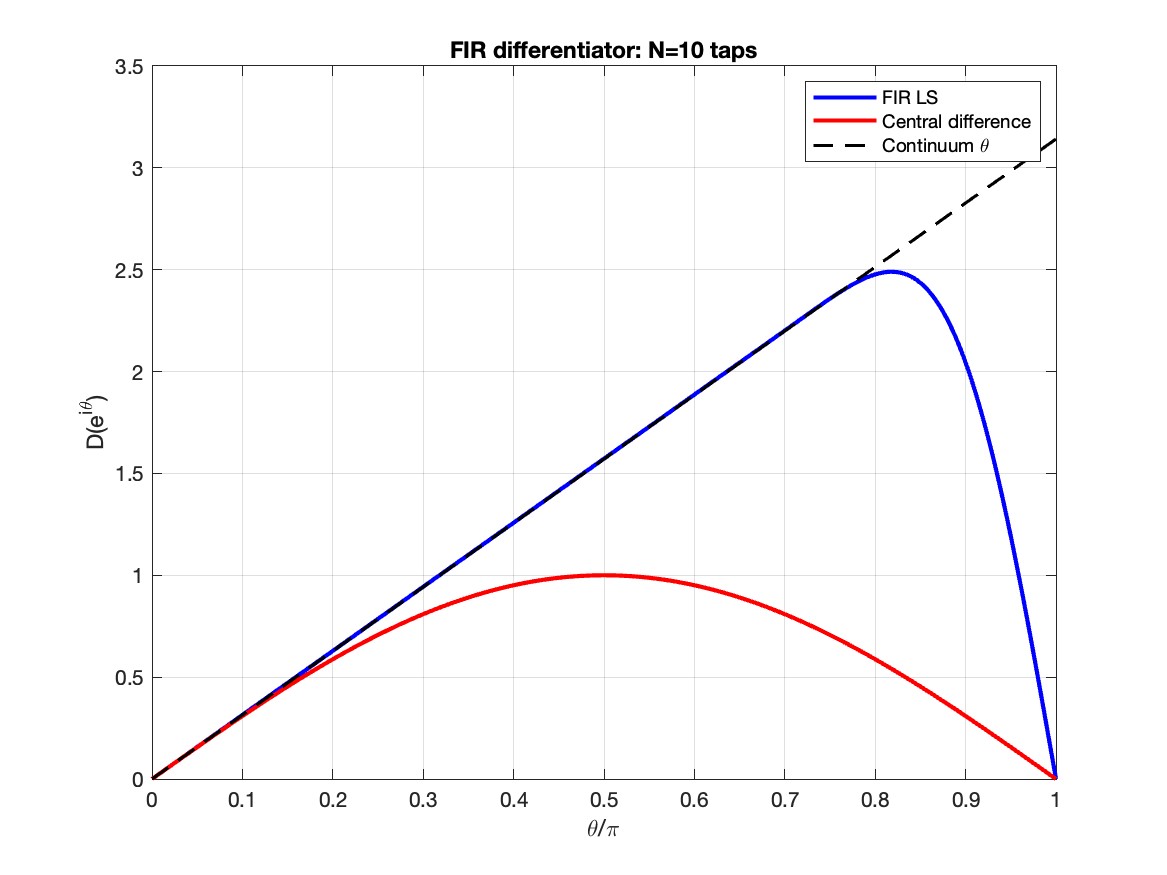} 
    \caption{The dispersion function  for $10$ taps.   }
    \label{kort}
\end{figure}

The accuracy of the FIR construction does not require a large
number of taps. Figure~\ref{kort} shows the dispersion relation obtained
with only $N=10$ taps, corresponding to a spatial stencil coupling
$2N+1=21$ lattice sites. Despite this short coupling range, the
FIR response follows the continuum momentum relation closely
through most of the physical Brillouin zone, with substantial
deviation occurring only at large momenta, approximately beyond
$0.8\pi$.

The character of the residual high-momentum sector is also
notably different from that of the central-difference operator.
Rather than forming an extended secondary branch with a
well-defined approximately linear dispersion, the FIR response
becomes strongly nonlinear as the Brillouin-zone boundary is
approached. Consequently, a finite-bandwidth wave packet centered
in this region experiences substantial dispersion because its
Fourier components acquire significantly different group
velocities. Only an increasingly narrow momentum distribution can
remain coherent near the residual zero.

This observation suggests that fermion doubling need not be
addressed solely by removing the additional zero of the momentum
operator. An alternative is to restrict the spectral region in
which that zero can support coherent, particle-like propagation.
The present work does not attempt to establish this as a general
criterion for eliminating fermion doubling, but the numerical
results indicate that finite-range FIR operators can strongly
suppress the propagation of localized excitations in the residual
high-momentum sector even for relatively short stencils.

The corresponding time-domain behavior was also examined by
launching finite-width free-electron wave packets at different
central momenta. For the $N=10$ operator, packets with central
momentum below approximately $0.8\pi$ propagate coherently, while
packets centered at higher momenta undergo pronounced dispersion.
This behavior is consistent with the strong curvature of the FIR
dispersion relation in this region: the different Fourier components
of a finite-bandwidth packet acquire substantially different group
velocities and therefore separate during propagation. Thus, while
the FIR operator retains a mathematical zero at the Brillouin-zone
boundary, the associated high-momentum region does not readily
support localized, coherently propagating wave packets.

\section{Relation to Existing Work}

The interpretation of fermion doubling as a consequence of spectral
replication has occasionally been noted in the lattice field theory
literature. The present work differs in adopting the Z-transform as
the primary mathematical framework rather than merely as an analogy.

The Wilson formulation \cite{Wilson}, staggered fermions, domain-wall
fermions, and overlap fermions \cite{Neuberger1998} all seek improved
lattice fermion operators within conventional lattice field theory.
The Neuberger overlap operator \cite{Neuberger1998} achieves an exact
lattice realization of chiral symmetry by satisfying the
Ginsparg--Wilson relation \cite{GW},
\begin{equation}
    D_{\mathrm{ov}}
    =
    \frac{1}{a}
    \left(
        1+\frac{A}{\sqrt{A^\dagger A}}
    \right),
\end{equation}
where $A$ is a shifted Wilson operator. Although translation invariant
on a periodic lattice, the corresponding position-space kernel is
exponentially local rather than strictly finite range. Moreover, the
operator is not normally applied through an explicit convolution.
Each application requires evaluating the matrix sign function,
typically through rational or polynomial approximations, which
entails repeated applications of the underlying sparse operator.
The resulting computational cost is substantially greater than that
of a conventional nearest-neighbour discretization.

A different approach to avoiding fermion doubling is provided by the
SLAC derivative \cite{Drell}, which constructs the lattice
derivative directly in momentum space. On a periodic lattice its
momentum-space representation reproduces the continuum momentum
throughout the Brillouin zone, thereby avoiding the additional zero
associated with the central-difference derivative. This spectral
accuracy comes at the cost of infinite spatial range: the corresponding
position-space kernel has a slowly decaying tail, so that every lattice
site is coupled to every other site. The nonlocal structure becomes
particularly significant when the derivative is coupled to gauge
fields, where gauge transport between arbitrarily separated sites is
required. The implications of this nonlocality for interacting gauge
theories have been discussed extensively in the literature.

The FIR construction proposed here follows a different design
philosophy. Rather than enforcing an exact lattice chiral symmetry
through the Ginsparg--Wilson relation, or reproducing the continuum
dispersion exactly at the cost of infinite spatial range as in the
SLAC construction, the momentum operator is obtained as the solution
of a constrained approximation problem. Among all finite-range
antisymmetric convolution operators of a given length, the
least-squares procedure constructs the operator whose frequency
response most closely approximates the continuum momentum operator
over the Brillouin zone. The resulting operator possesses a strictly
finite convolution kernel of length $2N+1$, whose coefficients are
computed once and then applied uniformly across the lattice.

This finite-range requirement is a central design principle of the
present work. In contrast to an infinitely extended derivative, a
finite-range operator requires gauge transport only over a bounded
number of lattice spacings when coupled to gauge fields. We do not
investigate the resulting interacting gauge theory in the present
work, and therefore make no claim that finite range alone resolves
the difficulties associated with nonlocal fermion formulations.
Rather, finite spatial range is taken as a practical constraint that
may make such an extension more tractable. This possibility needs to be studied, and is future work.  

Ghost suppression is achieved not by introducing a Wilson mass term,
but by constructing an operator whose spectral response closely
approximates the continuum derivative over the Brillouin zone. The
symmetry-enforced zero at the Brillouin-zone boundary is therefore
not removed. Instead, as the number of taps increases, the spectral
region surrounding this zero becomes increasingly narrow, while the
dispersion away from the zero approaches the continuum result.
Numerical propagation experiments show that wave packets with finite
momentum bandwidth in this residual ghost region undergo strong
dispersion, whereas coherent propagation requires an increasingly
narrow momentum distribution. At the same time, the proposed FIR
operators reproduce the continuum group velocity over the physical
momentum range substantially more accurately than the central-difference and Wilson discretizations considered here.

\section{Conclusions}

The paper presented a Z-transform framework for the analysis and
synthesis of lattice momentum operators. Within this framework,
translation-invariant lattice derivatives are represented by their
transfer functions on the unit circle, allowing the construction of
momentum operators to be viewed as a problem in constrained frequency-domain
approximation.

The formulation provides a new interpretation of the fermion doubling
problem. Rather than viewing doubling solely as a consequence of lattice
discretization, it is identified with the appearance of unwanted zeros
of the lattice momentum operator on the unit circle. This establishes a
direct connection between lattice field theory and the theory of digital
differentiators, where the placement of poles and zeros determines the
spectral properties of the operator.

The proposed framework naturally leads to a constructive design
procedure. By restricting attention to finite, antisymmetric convolution
kernels, the momentum operator is obtained as the solution of a
least-squares approximation problem in the frequency domain. The
resulting FIR operators preserve translation invariance and
anti-Hermiticity exactly, while systematically improving the
approximation to the continuum momentum operator as the stencil width is
increased.

Analytical arguments together with numerical simulations of free
electron propagation in one spatial dimension demonstrate that the
resulting operators reproduce the continuum dispersion relation and
group velocity substantially more accurately than the conventional
central-difference and Wilson discretizations over the physical momentum
range. In particular, ghost propagation is strongly suppressed because
the approximation error is concentrated into an increasingly narrow
neighbourhood of the Brillouin-zone boundary as the number of taps is
increased.

Unlike the overlap formulation, which is derived by enforcing the
Ginsparg--Wilson relation, the present construction is obtained by
solving a constrained approximation problem. The resulting operators
have a strictly finite convolution kernel whose coefficients are
precomputed once and then applied uniformly across the lattice.

Whether the proposed operators provide practical advantages in
interacting lattice gauge theories remains an open question. Their
performance in higher dimensions, in the presence of gauge fields, and
their relationship to established lattice fermion formulations require
further investigation. More broadly, the present work suggests that
digital filter design and approximation theory provide a useful
mathematical framework for the systematic construction and analysis of
lattice momentum operators.

A  promising direction for future work is the extension of
the finite-range FIR construction to gauge interactions.  The numerical
results presented here indicate that relatively short operators can
provide accurate free-particle propagation while suppressing coherent
propagation of localized excitations in the residual high-momentum
sector.   A gauge-covariant
extension would therefore require parallel transport over a finite
number of gauge links, in contrast to the infinite-range coupling of
the SLAC derivative.  This finite transporter range may make the FIR
construction substantially more amenable to gauge-covariant
implementation.  Whether the favourable spectral and propagation
properties demonstrated here survive in the presence of gauge fields
remains an open question and will be investigated in future work.


\section{Appendix}
\subsection{Proof of Impossibility of a Rational Momentum Operator}

\begin{proof}

Following Sanders \cite{Sanders2026}, we reformulate the existence 
problem by mapping the unit circle to the real axis via the 
M\"{o}bius transformation. Hence $E$ maps the circle to the real axis rather than the imaginary:
\begin{enumerate}
    \item $E(e^{i\theta}) = f(\theta) \in \mathbb{R}\ \forall\, 
    \theta \in [-\pi, \pi]$, \quad $f(-\theta) = -f(\theta)$
    \item $E(z) = 0$ on $|z| = 1 \iff z = 1$
    \item $E(e^{i\theta}) = \theta + O(\theta^2)$ \quad as $\theta \to 0$
    \item $E(z)$ is a rational function of $z$
\end{enumerate}

Note that $g(z) = -2i\dfrac{z-1}{z+1}$ satisfies these requirements. Note also that the inverse of $g$ is $\mathrm{inv}_g = \dfrac{2i - z}{z + 2i}$ and is  rational.

$E$ satisfies these criteria if and only if $F = E \circ \mathrm{inv}_g$ satisfies these requirements:
\begin{enumerate}
    \item $F(w) \in \mathbb{R}\ \forall\, w \in \mathbb{R}$, \quad 
    $F(-w) = -F(w)$
    \item $F(w) = 0 \iff w = 0$ \quad (on $\mathbb{R} \cup \{\infty\}$), 
    with $\lim_{w \to \infty} F(w) \neq 0$
    \item $F(w) = w + O(w^2)$ \quad as $w \to 0$
    \item $F(w)$ is a rational function of $w$
\end{enumerate}

A necessary condition is that $F(w) = wR(w^2)$, where $R$ is a real 
rational function of degree $\geq 0$.

The additional demand that $E$ should have no poles on the circle is 
the same as the demand that $F$ should have no poles on the real line 
and that $F$ should not diverge at infinity. This cannot be achieved, 
because the degree of $F$ is a positive odd number.

\end{proof}


\begin{thebibliography}{99}

\bibitem{Wilson}
K.~G.~Wilson,
``Confinement of Quarks,''
\emph{Physical Review D},
\textbf{10},
2445--2459,
1974.

\bibitem{Drell} 
S.D. Drell, M. Weinstein and S. Yankielowicz, \emph{Strong-coupling field theories. II. Fermions and gauge fields on a lattice}, Physical Review D, volume 14, number 6, pages 1627--1647, 1976.

\bibitem{Duitsers} J. Lenz, L. Pannullo, M. Wagner, B. Wellegehausen  and A. Wipf, \emph{Inhomogeneous phases in the Gross-Neveu model in
               1 + 1 dimensions at finite number of flavors}, Physical Review D,
    volume 101,
    number 9, 2020. 


\bibitem{Nielsen1}
H.~B.~Nielsen and M.~Ninomiya,
``No-Go Theorem for Regularizing Chiral Fermions,''
\emph{Physics Letters B},
\textbf{105},
219--223,
1981.

\bibitem{Nielsen2}
H.~B.~Nielsen and M.~Ninomiya,
``Absence of Neutrinos on a Lattice. I. Proof by Homotopy Theory,''
\emph{Nuclear Physics B},
\textbf{185},
20--40,
1981.


\bibitem{Montvay}
I.~Montvay and G.~M\"unster,
\emph{Quantum Fields on a Lattice},
Cambridge University Press,
Cambridge,
1994.

\bibitem{Oppenheim}
A.~V.~Oppenheim and R.~W.~Schafer,
\emph{Discrete-Time Signal Processing},
3rd Edition,
Pearson,
2010.




\bibitem{Neuberger1998}
H. Neuberger, \emph{Exactly massless quarks on the lattice}, Physics Letters B, volume 417, pages 141-144, 1998. 


 \bibitem{Weinberg}
S.~Weinberg,
\emph{The Quantum Theory of Fields},
Volume~I,
Cambridge University Press,
Cambridge,
1995.

\bibitem{Karsten} 
L.H. Karsten, and J. Smit, \emph{The vacuum polarization with SLAC lattice fermions}, Physics Letters B, Vol.  85,
 pages 100--102, 1979. 



\bibitem{GW}
P.~H.~Ginsparg and K.~G.~Wilson,
``A Remnant of Chiral Symmetry on the Lattice,''
\emph{Physical Review D},
\textbf{25},
2649--2657,
1982.


\bibitem{Sanders2026}
C.~Sanders, private communication, 2026.



\end{thebibliography}
\end{document}